\documentclass[journal,11pt,onecolumn,draftclsnofoot]{IEEEtran}
\usepackage{amsmath,amsfonts}
\usepackage{algorithmic}
\usepackage{algorithm}
\usepackage{array}
\usepackage[caption=false,font=footnotesize,labelfont=rm,textfont=rm]{subfig}
\usepackage{textcomp}
\usepackage{stfloats}
\usepackage{url}
\usepackage{verbatim}
\usepackage[dvipdfmx]{graphicx}
\usepackage{cite}
\usepackage{bm,mathrsfs}
\usepackage{color}
\usepackage{siunitx}

\DeclareMathOperator*{\argmin}{arg\,min}

\DeclareMathOperator*{\minimize}{minimize}
\usepackage{xcolor}

\colorlet{commentColor}{blue}

\begin{document}

\title{Physics-Informed Machine Learning\\For Sound Field Estimation\\
\large{Invited paper for the IEEE SPM Special Issue on Model-based Data-Driven Audio Signal Processing}}

\author{Shoichi Koyama,~\IEEEmembership{Senior Member,~IEEE,}
Juliano G. C. Ribeiro,~\IEEEmembership{Member,~IEEE,}
Tomohiko Nakamura,~\IEEEmembership{Member,~IEEE,}
Natsuki Ueno,~\IEEEmembership{Member,~IEEE,}
and Mirco Pezzoli,~\IEEEmembership{Member,~IEEE}
        % <-this % stops a space
\thanks{This work was supported by JST FOREST Program, Grant Number JPMJFR216M, JSPS KAKENHI, Grant Number 23K24864, and the European Union under the Italian National Recovery and Resilience Plan (NRRP) of NextGenerationEU, partnership on ``Telecommunications of the Future'' (PE00000001---program ``RESTART'').}% <-this % stops a space
\thanks{Manuscript received April xx, 20xx; revised August xx, 20xx.}}

\markboth{IEEE Signal Processing Magazine,~Vol.~xx, No.~x, August~20xx}%
{Shell \MakeLowercase{\textit{et al.}}: A Sample Article Using IEEEtran.cls for IEEE Journals}

\IEEEpubid{0000--0000/00\$00.00~\copyright~2024 IEEE}

\maketitle
\begin{abstract} 
The area of study concerning the estimation of spatial sound, i.e., the distribution of a physical quantity of sound such as acoustic pressure, is called \textit{sound field estimation}, which is the basis for various applied technologies related to spatial audio processing. The sound field estimation problem is formulated as a function interpolation problem in machine learning in a simplified scenario. However, high estimation performance cannot be expected by simply applying general interpolation techniques that rely only on data. The physical properties of sound fields are useful a priori information, and it is considered extremely important to incorporate them into the estimation. In this article, we introduce the fundamentals of \textit{physics-informed machine learning (PIML)} for sound field estimation and overview current PIML-based sound field estimation methods.
\end{abstract}


\section{Introduction}
\label{sec:intro}

\begin{figure}[!t]
\centering
\includegraphics[width=3.8in,clip]{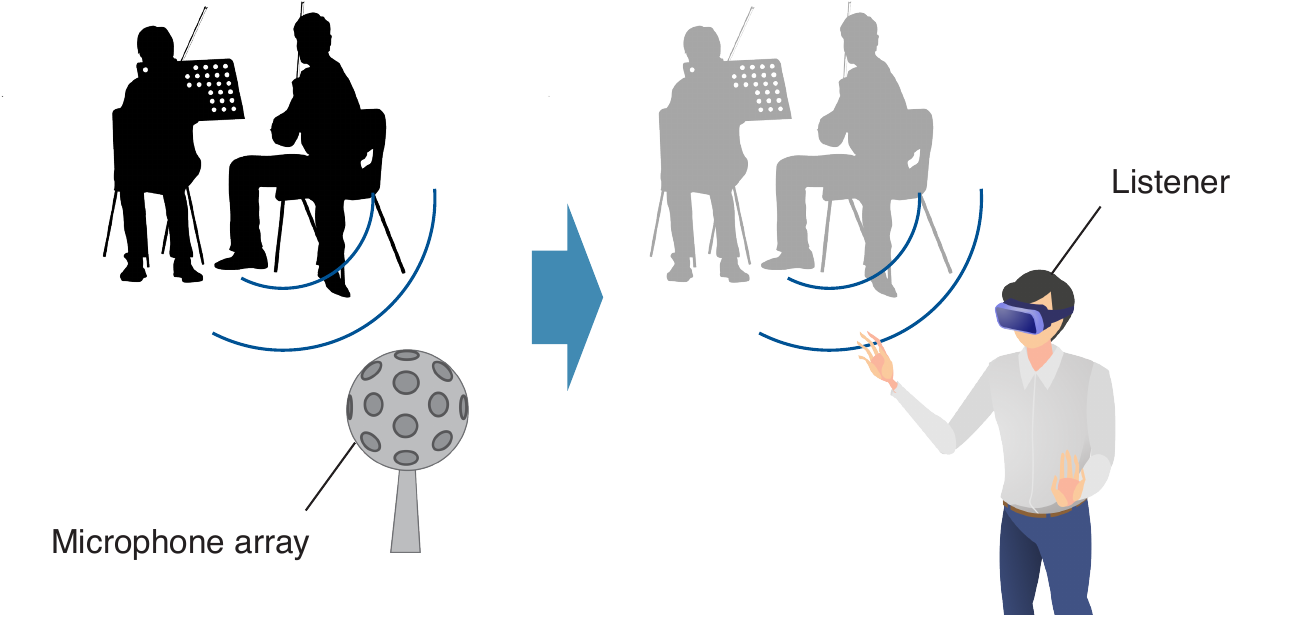}
\caption{Sound field recording for VR audio. The sound field captured by a microphone array is reproduced by loudspeakers or headphones. The estimated sound field should account for listener movement and rotation within large regions.}
\label{fig:vraudio}
\end{figure}

Sound field estimation, which is also referred to as sound field reconstruction, capturing, and interpolation, is a fundamental problem in acoustic signal processing, which is aimed at reconstructing a spatial acoustic field from a discrete set of microphone measurements. This essential technology has a wide variety of applications, such as room acoustic analysis, the visualization/auralization of an acoustic field, spatial audio reproduction using a loudspeaker array or headphones, and active noise cancellation in a spatial region. In particular, virtual/augmented reality (VR/AR) audio would be one of the most remarkable recent applications of this technology, as it requires capturing a sound field in a large region by multiple microphones (see Fig.~\ref{fig:vraudio}). The spatial reconstruction of room impulse responses (RIRs) or acoustic transfer functions (ATFs), which is a special case of sound field estimation, can be applied to the estimation of steering vectors for beamforming and also head-related transfer functions (HRTFs) for binaural reproduction.

Sound field estimation has been studied for a number of years. Estimating a sound field in the inverse direction of wave propagation based on wave domain (or spatial frequency domain) processing, i.e., the basis expansion of a sound field into plane wave or spherical wave functions, has been particularly applied to acoustic imaging tasks~\cite{Williams:FourierAcoust}. This technique has been introduced in the signal-processing field in recent decades. In particular, spherical harmonic domain processing using a spherical microphone array has been intensively investigated because its isotropic nature is suitable for spatial sound processing~\cite{Rafaely:FundSphArrayProc}. Wave domain processing has been applied to, for instance, spatial audio recording, source localization, source separation, and spatial active noise control. 

The theoretical foundation of wave domain processing is derived from expansion representations by the solutions of the governing partial differential equations (PDEs) of acoustic fields, i.e., the wave equation in the time domain and the Helmholtz equation in the frequency domain. Since the basis functions used in wave domain processing, such as plane wave and spherical wave functions, are solutions of these PDEs in a region not including sound sources or scattering objects, the estimate is guaranteed to satisfy the wave and Helmholtz equations. 

Although the classical methods are solely based on physics, advanced signal processing and machine learning techniques have also been applied to the sound field estimation problem. In particular, sparse optimization techniques have been investigated to improve the estimation accuracy while preserving the physical constraints for the estimate~\cite{Bertin:CSAbook2015,Antonello:IEEE_ACM_J_ASLP2017,Murata:IEEE_J_SP2018}. The physical constraints are usually imposed by the expansion representation of the sound field as used in the classical methods.  

On the basis of great successes in various fields, (deep) neural networks~\cite{Goodfellow:DL} have also been incorporated into sound field estimation techniques in recent years to exploit their high representational power. The network weights are optimized to predict the sound field by using a set of training data. Basically, this approach is purely data-driven; that is, prior information on physical properties is not taken into account, as opposed to the methods mentioned above. To induce the estimate to satisfy the governing PDEs, the loss function for evaluating the deviation from the governing PDEs has been incorporated, and such a technique is referred to as the \textit{physics-informed neural networks (PINNs)}~\cite{Karniadakis:NatRevPhus2021,Raissi:CompPhys2019}, which have been applied to various forward and inverse problems for physical fields, such as flow field~\cite{Cuomo:JSciComp2022}, seismic field~\cite{Lin:IEEE_SPM2023}, and acoustic field~\cite{Borrel-Jensen:PNAS2023}. PINNs help prevent unnatural distortions in the estimate, which are unlikely to occur in practical acoustic fields, particularly when the number of observations is small. Currently, there exist various methods using PINNs or other neural networks for sound field estimation as described in detail in the later sections~\cite{Chen:APSIPA2023,Ma:arxiv2024,Olivieri:EURASIP2024,Karakonstantis:JASA2024}. 

Physical properties are useful prior information in sound field estimation and methods based on physical properties have many advantages over purely data-driven approaches. Although sound field estimation methods have historically used physical constraints explicitly, with the recent development of machine learning techniques, \textit{physics-informed machine learning (PIML)} in sound field estimation continues to grow significantly, especially after the advent of PINN~\cite{Oliveri:Sensors2021,Shigemi:IWAENC2022,Karakonstantis:JASA2023,Ribeiro:ICASSP2023}. We describe how to embed physical properties for various machine learning techniques, such as linear regression, kernel regression, and neural networks, in the sound field estimation problem and summarize the current PIML-based methods for sound field estimation. Furthermore, current limitations and future outlooks are also discussed.

\section{Sound field estimation problem}
\label{sec:problem}

\begin{figure*}[!t]
\centering
\subfloat[Interior problem]{\includegraphics[width=2.0in,clip]{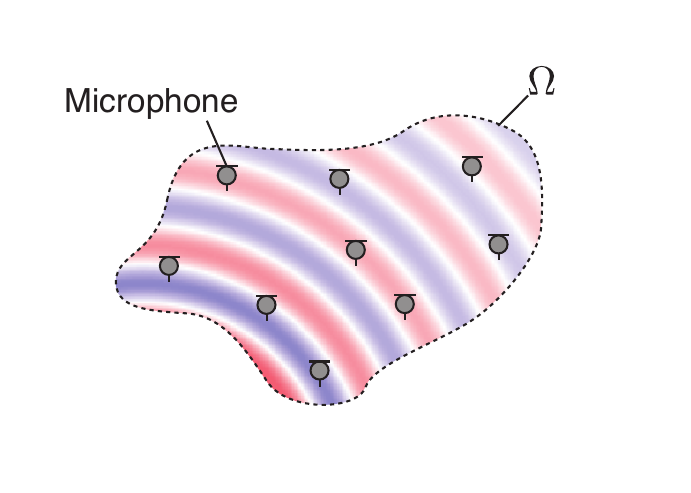}}
\hspace{40pt}
\subfloat[Exterior problem]{\includegraphics[width=2.0in,clip]{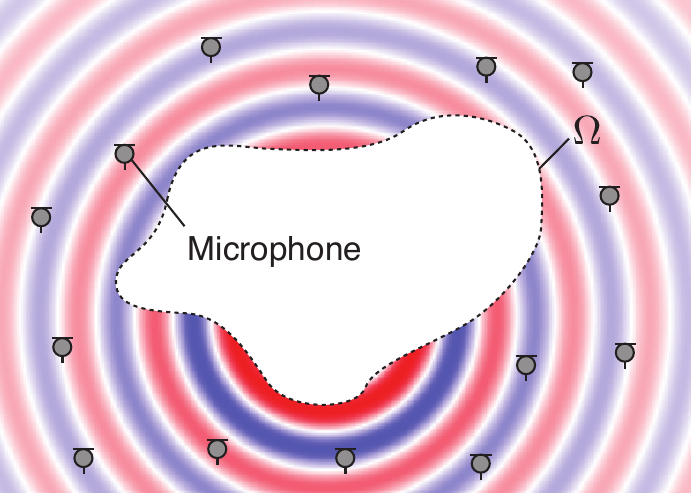}}
\caption{Sound field estimation of interior and exterior regions. In the interior problem, the sound field in the source-free interior region $\Omega$ is estimated. In contrast, the sound field in the source-free exterior region $\mathbb{R}^3\backslash\Omega$ is estimated in the exterior problem.}
\label{fig:sf_est}
\end{figure*}

First, we formulate the sound field estimation problem. In this article, we mainly focus on an interior problem, that is, estimating a sound field inside a source-free region of the target with multiple microphones (see Fig.~\ref{fig:sf_est}). Several methods introduced in this article are also applicable to the exterior or radiation problem. 

The target region in the three-dimensional space is denoted as $\Omega \subset \mathbb{R}^3$. The pressure distribution in $\Omega$ is represented by $U: \Omega \times \mathbb{R} \rightarrow \mathbb{R}$ in the time domain and $u: \Omega \times \mathbb{R} \rightarrow \mathbb{C}$ in the frequency domain, which is generated by sound waves propagating from one or more sound sources outside $\Omega$ and also by reverberation. Here, $U$ and $u$ are continuous functions of the position $\bm{r}\in\Omega$ and the time $t\in\mathbb{R}$ or the angular frequency $\omega\in\mathbb{R}$. $M$ microphones are placed at arbitrary positions inside $\Omega$. The objective of the sound field estimation problem is to reconstruct $U$ or $u$ from microphone observations. 

In the frequency domain, the estimation can be performed at each frequency assuming signal independence between frequencies. If the estimator of $u$ is linear in the frequency domain, a finite-impulse-response (FIR) filter for estimating a sound field generated by usual sounds, such as speech and music, can be designed. Otherwise, the estimation is performed in the time--frequency domain, for example, by the short-time Fourier transform (STFT), or in the time domain. Particularly when the estimator is adapted to the environment, the direct estimation in the time domain is generally more challenging because the estimator for $U$ may have a large dimension of spatiotemporal functions. When the source signal is an impulse signal, which means that $U$ is a spatiotemporal function of an RIR, the problem can be relatively tractable because the time samples to be estimated are reduced to the length of RIRs at most and the estimation is generally performed offline.

For simplicity, the microphones are assumed to be omnidirectional, i.e., pressure microphones. In this scenario, the sound field estimation problem is equivalent to the general interpolation problem because the observations are spatial (or spatiotemporal) samples of the multivariate scalar function to be estimated. Specifically, when the microphone observations in the frequency domain are denoted by $\bm{s}(\omega)=[s_1(\omega), \ldots, s_M(\omega)]^{\mathsf{T}}\in\mathbb{C}^{M}$ ($(\cdot)^{\mathsf{T}}$ is the transpose) obtained by the pressure microphones at $\{\bm{r}_m\}_{m=1}^M$, the observed values of $\{s_m(\omega)\}_{m=1}^M$ can be regarded as $\{u(\bm{r}_m, \omega)\}_{m=1}^M$ contaminated by some noise. In the time domain, the sound field should be estimated from the observations of $M$ microphones and $T$ time samples. Most methods introduced in this article can be generalized to the case of directional microphones because the measurement process can be assumed to be a linear operation. It is also possible for most methods to be generalized to estimate expansion coefficients of spherical wave functions of the sound field at a predetermined expansion center, which is particularly useful for spatial audio applications such as \textit{ambisonics}~\cite{Zotter:Ambisonics}. 

We start with techniques of embedding physical properties in general interpolation techniques. Then, current studies of sound field estimation based on PIML are introduced.

\section{Embedding physical properties in interpolation techniques}
\label{sec:emb-phys}

In general interpolation techniques, the estimation of the continuous function $f: \mathbb{R}^P \to \mathbb{K}$ ($\mathbb{K}$ is $\mathbb{R}$ or $\mathbb{C}$, i.e., $f$ is $U$ or $u$) from a discrete set of observations $\bm{y}\in\mathbb{K}^I$ at the sampling points $\{\bm{x}_i\}_{i=1}^I$ is achieved by representing $f$ with some model parameters $\bm{\theta}$ and solving the following optimization problem:
\begin{align}
 \minimize_{\bm{\theta}} \mathcal{L}\left( \bm{y}, \bm{f}(\{\bm{x}_i\}_{i=1}^I;\bm{\theta}) \right) + \mathcal{R}(\bm{\theta}),
\label{eq:opt}
\end{align}
where $\bm{f}(\{\bm{x}_i\}_{i=1}^I; \bm{\theta}) = [f(\bm{x}_1; \bm{\theta}), \ldots, f(\bm{x}_I; \bm{\theta})]^{\mathsf{T}}\in\mathbb{K}^I$ is the vector of the discretized function $f$ represented by $\bm{\theta}$, $\mathcal{L}$ is the loss function for evaluating the distance between $\bm{y}$ and $f$ at $\{\bm{x}_i\}_{i=1}^I$, and $\mathcal{R}$ is the regularization term for $\bm{\theta}$ to prevent overfitting. A typical loss function is the squared error between $\bm{y}$ and $\bm{f}$, which is written as
\begin{align}
 \mathcal{L}\left( \bm{y}, \bm{f}(\{\bm{x}_i\}_{i=1}^I;\bm{\theta}) \right) = \| \bm{y} - \bm{f}(\{\bm{x}_i\}_{i=1}^I; \bm{\theta}) \|^2.
 \label{eq:loss_l2}
\end{align}
One of the simplest regularization terms is the squared $\ell_2$-norm penalty (Tikhonov regularization) for $\bm{\theta}$ defined as
\begin{align}
 \mathcal{R}(\bm{\theta}) = \lambda \|\bm{\theta}\|^2,
 \label{eq:reg_l2}
\end{align}
where $\lambda$ is the positive constant parameter. By using the solution of \eqref{eq:opt}, which is denoted as $\hat{\bm{\theta}}$, we can estimate the function $f$ as $f(\bm{x}; \hat{\bm{\theta}})$.

The general interpolation techniques described above highly depend on data. Prior information on the function to be estimated is basically useful for preventing overfitting, whose simplest example is the squared $\ell_2$-norm penalty \eqref{eq:reg_l2}. In the context of sound field estimation, one of the most beneficial lines of information will be the physical properties of the sound field. In particular, a governing PDE of the sound field, i.e., the wave equation in the time domain or the Helmholtz equation in the frequency domain, is an informative property because the function to be estimated should satisfy the governing PDE. The homogeneous wave and Helmholtz equations are respectively written as
\begin{align}
 \left( \nabla_{\bm{r}}^2 - \frac{1}{c^2}\frac{\partial^2}{\partial t^2}  \right) U(\bm{r},t) = 0 
 \end{align}
 and
 \begin{align}
 \left( \nabla_{\bm{r}}^2 + k^2 \right) u(\bm{r},\omega) = 0,
\end{align}
where $c$ is the sound speed and $k=\omega/c$ is the wave number. We introduce several techniques to embed these physical properties in the interpolation.

\subsection{Regression with basis expansion into element solutions of governing PDEs}
\label{sec:basis-exp}

A widely used model for $f$ is a linear combination of a finite number of basis functions. We define the basis functions and their weights as $\{\varphi_l(\bm{x})\}_{l=1}^L$ and $\{\gamma_l\}_{l=1}^L$, respectively, with $\varphi_l: \mathbb{R}^P \to \mathbb{K}$ and $\gamma_l \in \mathbb{K}$. Then, $f$ is expressed as
\begin{align}
 f(\bm{x}; \bm{\gamma}) &= \sum_{l=1}^L \gamma_l \varphi_l(\bm{x}) \notag\\
&= \bm{\varphi}(\bm{x})^{\mathsf{T}} \bm{\gamma},
\label{eq:basis_exp}
\end{align}
where $\bm{\varphi}(\bm{x})=[\varphi_1,\ldots,\varphi_L]^{\mathsf{T}} \in \mathbb{K}^L$ and $\bm{\gamma}=[\gamma_1,\ldots,\gamma_L]^{\mathsf{T}} \in \mathbb{K}^L$. Thus, $f$ is interpolated by estimating $\bm{\gamma}$ from $\bm{y}$.

\begin{figure*}[!t]
\centering
\subfloat[Plane wave functions]{\includegraphics[width=6.2in,clip]{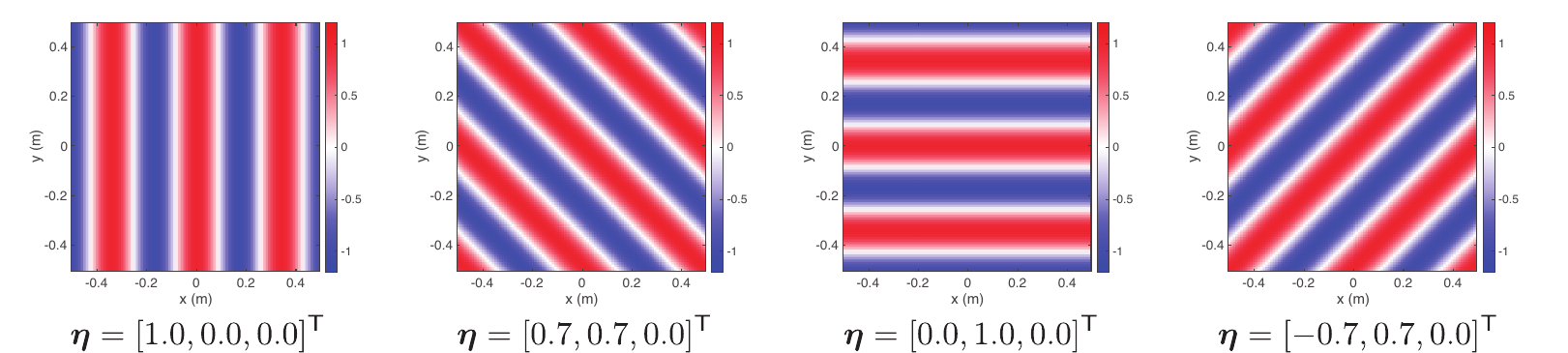}}\\
\subfloat[Spherical wave functions]{\includegraphics[width=6.2in,clip]{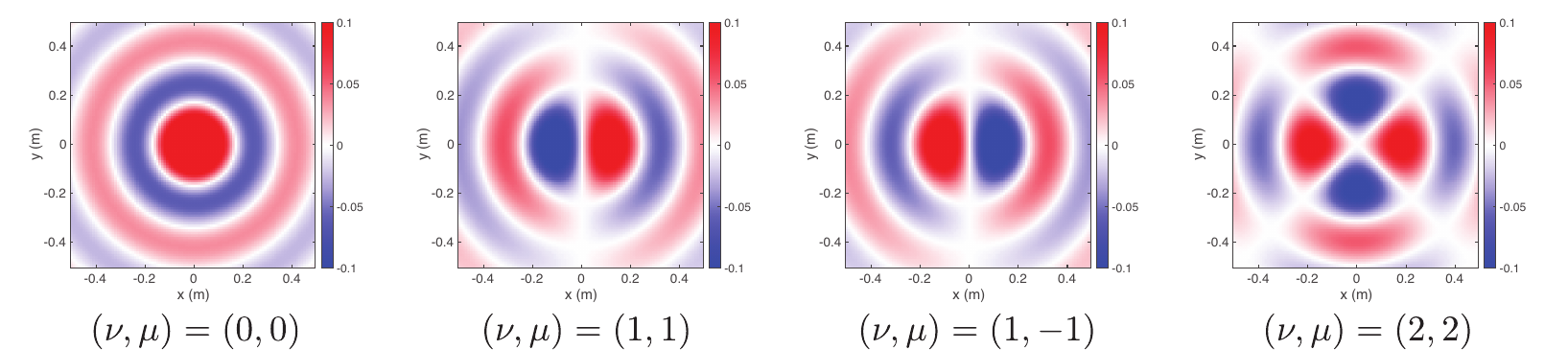}}
\caption{Basis functions used to constrain the solution to satisfy the Helmholtz equation on the $x$--$y$ plane at the frequency of $1000~\mathrm{Hz}$. (a) Plane wave functions. (b) Spherical wave functions. Real parts of the functions are plotted.}
\label{fig:basis_fun}
\end{figure*}

A well-established technique used to constrain the estimate $f$ to the governing PDEs is the expansion into the element solutions of PDEs, which is particularly investigated in the frequency domain to constrain the estimate to satisfy the Helmholtz equation. Several expansion representations for the Helmholtz equation have been applied to the sound field estimation problem. When considering the interior problem, the following representations are frequently used~\cite{Colton:InvAcoust_2013}:
\begin{itemize}
 \item Plane wave expansion (or Herglotz wave function)
\begin{align}
 u(\bm{r},\omega) = \int_{\mathbb{S}_2} \tilde{u}(\bm{\eta},\omega) \mathrm{e}^{-\mathrm{j} k \langle \bm{\eta}, \bm{r} \rangle} \mathrm{d} \bm{\eta},
\label{eq:pw-exp}
\end{align}
where $\tilde{u}$ is the complex amplitude of plane waves arriving from the direction $\bm{\eta} \in\mathbb{S}_2$ ($\mathbb{S}_2$ is the unit sphere). Although the plane wave function is usually defined by using the propagation direction $\tilde{\bm{\eta}}=-\bm{\eta}$, \eqref{eq:pw-exp} is defined by using the arrival direction $\bm{\eta}$ for simplicity in the later formulations.
\item Spherical wave function expansion
\begin{align}
 u(\bm{r},\omega) = \sum_{\nu=0}^{\infty} \sum_{\mu=-\nu}^{\nu} \mathring{u}_{\nu,\mu}(\bm{r}_{\mathrm{o}},\omega) j_{\nu}(k\|\bm{r} - \bm{r}_{\mathrm{o}}\|) Y_{\nu,\mu} \left( \frac{\bm{r}-\bm{r}_{\mathrm{o}}}{\|\bm{r}-\bm{r}_{\mathrm{o}}\|} \right),
\label{eq:swf-exp}
\end{align}
where $\mathring{u}_{\nu,\mu}$ is the expansion coefficient of the order $\nu$ and degree $\mu$, $\bm{r}_{\mathrm{o}}$ is the expansion center, $j_{\nu}$ is the $\nu$th-order spherical Bessel function, and $Y_{\nu, \mu}$ is the spherical harmonic function of the order $\nu$ and degree $\mu$, accepting as arguments the azimuth and zenith angles of the unit vector $(\bm{r} - \bm{r}_{\mathrm{o}})/\|\bm{r} - \bm{r}_{\mathrm{o}}\|$.
 \item Equivalent source distribution (or single-layer potential)
\begin{align}
 u(\bm{r}, \omega) = \int_{\partial \Omega} \breve{u}(\bm{r}^{\prime},\omega)\frac{\mathrm{e}^{\mathrm{j} k (\bm{r}-\bm{r}^{\prime})}}{4\pi \|\bm{r}-\bm{r}^{\prime}\|} \mathrm{d}\bm{r}^{\prime},
\label{eq:esd}
\end{align}
where $\breve{u}$ is the weight of the equivalent source, i.e., the point source, and $\partial \Omega$ is the boundary of $\Omega$. The boundary surface of the equivalent source is not necessarily identical to $\partial \Omega$ if it encloses the region $\Omega$.
\end{itemize}
Examples of the element solutions are plotted in Fig.~\ref{fig:basis_fun}. By discretizing the integration operation in \eqref{eq:pw-exp} and \eqref{eq:esd} or truncating the infinite-dimensional series expansion in \eqref{eq:swf-exp}, and setting the element solutions as $\{\varphi_l(\bm{x})\}_{l=1}^L$ and their weights as $\{\gamma_l\}_{l=1}^L$, we can approximate the above representations as the finite-dimensional basis expansions in \eqref{eq:basis_exp}. 

When the squared error loss function \eqref{eq:loss_l2} and $\ell_2$-norm penalty \eqref{eq:reg_l2} are used, \eqref{eq:basis_exp} becomes a simple least squares problem (or linear ridge regression). The closed-form solution of $\bm{\gamma}$ is obtained as
\begin{align}
 \hat{\bm{\gamma}} &= \argmin_{\bm{\gamma}\in\mathbb{C}^L} \|\bm{y} - \bm{\Phi} \bm{\gamma} \|^2 + \lambda \|\bm{\gamma}\|^2 \notag\\
&= \left( \bm{\Phi}^{\mathsf{H}} \bm{\Phi} + \lambda \bm{I} \right)^{-1} \bm{\Phi}^{\mathsf{H}} \bm{y},
\label{eq:linreg}
\end{align}
where $\bm{\Phi}=[\bm{\varphi}(\bm{x}_1), \ldots, \bm{\varphi}(\bm{x}_I)]^{\mathsf{T}} \in \mathbb{K}^{I \times L}$, $\bm{I}$ is the identity matrix, and $(\cdot)^{\mathsf{H}}$ is the conjugate transpose (equivalent to the transpose $(\cdot)^{\mathsf{T}}$ for a real-valued matrix). Then, $f$ is estimated using $\hat{\bm{\gamma}}$ as in \eqref{eq:basis_exp} under the constraint that $f$ satisfies the governing PDE.

Various regularization terms for $\mathcal{R}$ other than the $\ell_2$-norm penalty have been investigated to incorporate prior knowledge on the structure of $\bm{\gamma}$. For example, the $\ell_1$-norm penalty is widely used to promote sparsity on $\bm{\gamma}$ when $L \gg I$ in the context of sparse optimization and compressed sensing~\cite{Donoho:IEEE_J_IT2006}. However, iterative algorithms, such as the fast iterative shrinkage thresholding algorithm and iteratively reweighted least squares, must be used to obtain the solution~\cite{Elad:Sparse}.

When applying the linear ridge regression, how to decide the number of basis functions $L$, i.e., the number of plane waves \eqref{eq:pw-exp} or point sources \eqref{eq:esd} and truncation order in \eqref{eq:swf-exp}, is not a trivial task. For the spherical wave function expansion, using the $\Omega$ radius $R$, $\lceil kR \rceil$ and $\lceil \mathrm{e}kR/2 \rceil$ are empirically known to be an appropriate truncation order when the target region $\Omega$ is a sphere. The sparsity-promoting regularization can help find an admissible number of basis functions using a dictionary matrix ($\bm{\Phi}$ in \eqref{eq:linreg}) constructed using a large number of basis functions~\cite{Bertin:CSAbook2015,Antonello:IEEE_ACM_J_ASLP2017,Murata:IEEE_J_SP2018}. 

The finite-dimensional expansion into the element solutions of the governing PDE is a powerful and flexible technique to embed the physical properties, which has also been applied to the estimation of the exterior field and region including sources or scattering objects. This technique is also generalized to the cases of using directional microphones and estimating expansion coefficients of spherical wave functions~\cite{Laborie:AESconv2003,Poletti:J_AES_2005}.

\subsection{Kernel regression using reproducing kernel Hilbert space for governing PDE}
\label{sec:ker}

Kernel regression is classified as a nonparametric regression technique. The kernel function $\kappa: \mathbb{R}^P \times \mathbb{R}^P \to \mathbb{K}$ should be defined as a similarity function expressed as an inner product defined on some functional space $\mathscr{H}$, $\kappa(\bm{x}, \bm{x}^{\prime}) = \langle \bm{\varphi}(\bm{x}), \bm{\varphi}(\bm{x}^\prime) \rangle_{\mathscr{H}}$. Thus, $\bm{\varphi}$ can be infinite-dimensional, which is one of the major differences from the basis expansion described in Section~\ref{sec:basis-exp}. On the basis of the representer theorem~\cite{Murphy:ML}, $f$ is represented by the weighted sum of $\kappa$ at the sampling points as
\begin{align}
 f(\bm{x}; \bm{\alpha}) &= \sum_{i=1}^I \alpha_i \kappa(\bm{x},\bm{x}_i) \notag\\
&= \bm{\kappa}(\bm{x})^{\mathsf{T}}\bm{\alpha},
\label{eq:ker_rep}
\end{align}
where $\bm{\alpha}=[\alpha_1, \ldots, \alpha_I]^{\mathsf{T}} \in \mathbb{K}^I$ are the weight coefficients and $\bm{\kappa}(\bm{x})=[\kappa(\bm{x},\bm{x}_1), \ldots, \kappa(\bm{x},\bm{x}_I)]^{\mathsf{T}}$ is the vector of kernel functions. In the kernel ridge regression~\cite{Murphy:ML}, $\bm{\alpha}$ is obtained as
\begin{align}
 \hat{\bm{\alpha}} = (\bm{K} + \lambda \bm{I})^{-1} \bm{y},
\end{align}
with the Gram matrix $\bm{K} \in \mathbb{K}^{I \times I}$ defined as
\begin{align}
 \bm{K} = 
\begin{bmatrix}
 \kappa(\bm{x}_1,\bm{x}_1) & \cdots & \kappa(\bm{x}_1,\bm{x}_I) \\
 \vdots & \ddots & \vdots \\
 \kappa(\bm{x}_I,\bm{x}_1) & \cdots & \kappa(\bm{x}_I,\bm{x}_I) 
\end{bmatrix}.
\end{align}
Then, $f$ is interpolated by substituting $\hat{\bm{\alpha}}$ into \eqref{eq:ker_rep}.

\begin{figure*}[!t]
\centering
\includegraphics[width=6.0in,clip]{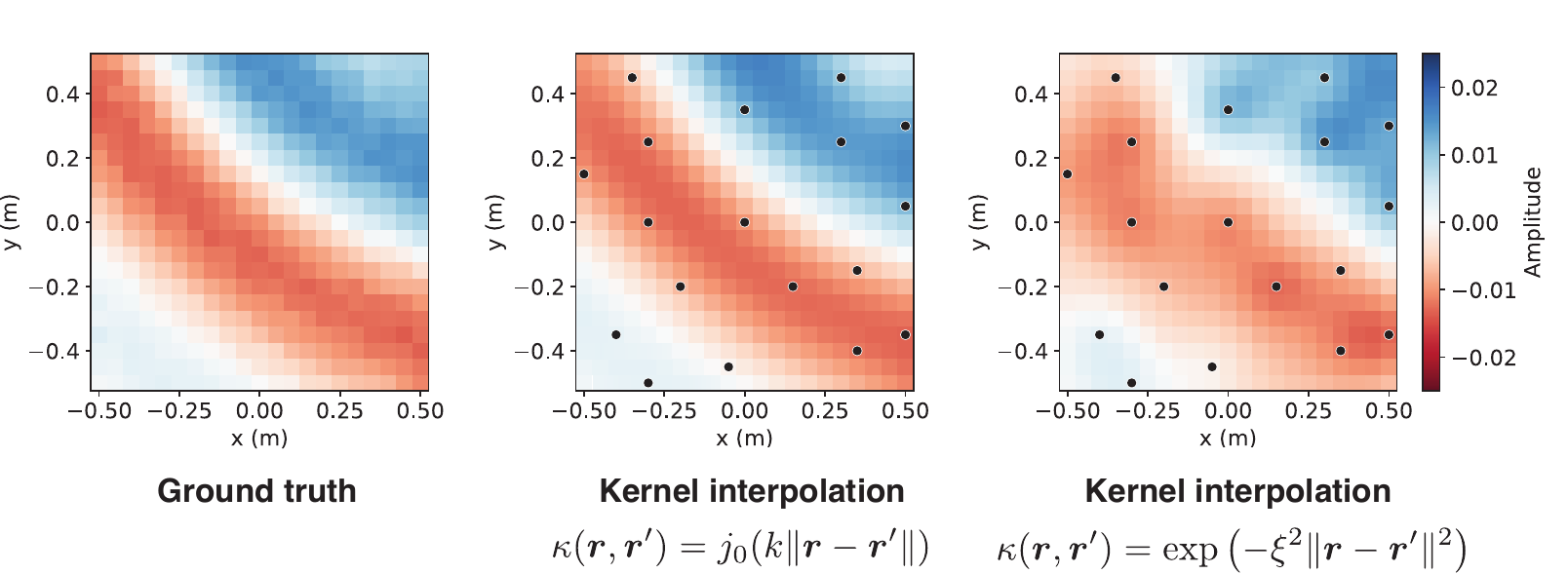}
\caption{Pressure distribution on the two-dimensional plane estimated by the kernel method using MeshRIR dataset~\cite{Koyama:WASPAA2021}. Black dots indicate the microphone positions ($M=18$). The kernel function of the spherical Bessel function \eqref{eq:ker} and Gaussian kernel ($\xi=1.0k$) are compared. The source is an ordinary loudspeaker, and the source signal is the pulse signal lowpass-filtered up to $500~\mathrm{Hz}$. The use of the physics-constrained kernel function improved the reconstruction accuracy. The figure is taken from \cite{Koyama:WASPAA2021}, with an additional comparison with Gaussian kernel.}
\label{fig:sf_est_kernel}
\end{figure*}

In the kernel regression, the reproducing kernel Hilbert space to seek a solution must be properly defined, which also defines the reproducing kernel function $\kappa$. In general machine learning, some assumptions for the data structure are imposed to define $\kappa$. For example, the Gaussian kernel function~\cite{Murphy:ML} is frequently used to induce the smoothness of the function to be interpolated. The reproducing kernel Hilbert space $\mathscr{H}$, as well as the kernel function $\kappa$, can be defined to constrain the solution to satisfy the Helmholtz equation~\cite{Ueno:IEEE_SPL2018}. The solution space of the homogeneous Helmholtz equation is defined by the plane wave expansion \eqref{eq:pw-exp} or spherical wave function expansion \eqref{eq:swf-exp} with square-integrable or square-summable expansion coefficients. We here define the inner product over $\mathscr{H}$ using the plane wave expansion with positive directional weighting $w: \mathbb{S}_2 \to \mathbb{R}_{\ge 0}$ as~\cite{Ueno:IEEE_J_SP2021}
\begin{align}
 \langle u_1, u_2 \rangle_{\mathscr{H}} = \frac{1}{4\pi} \int_{\mathbb{S}_2} \frac{1}{w(\bm{\eta})} \tilde{u}_1(\bm{\eta})^{\ast} \tilde{u}_2(\bm{\eta}) \mathrm{d}\bm{\eta}. \label{eq:ker-ip}
 %& \|u\|_{\mathscr{H}} = \sqrt{\langle u, u \rangle_{\mathscr{H}}}, \label{eq:ker-norm}
\end{align}
The directional weighting function $w$ is designed to incorporate prior knowledge of the directivity pattern of the target sound field by setting $\bm{\xi}$ in the approximate directions of sources or early reflections. We represent $w$ as a unimodal function using the von Mises--Fisher distribution~\cite{Mardia:DirStat} defined by
\begin{align}
 w(\bm{\eta}) = \frac{1}{C(\beta)} \mathrm{e}^{\beta \langle \bm{\eta},  \bm{\xi} \rangle}, 
 \label{eq:w_dir}
\end{align}
where $\bm{\xi}\in\mathbb{S}_2$ is the peak direction, $\beta$ is the parameter for controlling the sharpness of $w$, and $C(\beta)$ is the normalization factor
\begin{align}
 C(\beta) = 
\begin{cases}
 1, & \beta = 0 \\
 \frac{\mathrm{sinh}(\beta)}{\beta}, & \text{otherwise} 
\end{cases}.
\end{align}
Then, the kernel function is analytically expressed as
\begin{align}
 \kappa(\bm{r},\bm{r}^{\prime}) = \frac{1}{C(\beta)} j_0 \left( \sqrt{ \left(k(\bm{r}-\bm{r}^{\prime}) - \mathrm{j} \beta \bm{\xi}\right)^{\mathsf{T}} \left(k(\bm{r}-\bm{r}^{\prime}) - \mathrm{j} \beta \bm{\xi}\right) } \right).
\label{eq:ker-dir}
\end{align}
When the directional weighting $w$ is uniform, i.e., no prior information on the directivity, $w$ is set as $1$, and the kernel function is simplified as
\begin{align}
 \kappa(\bm{r},\bm{r}^{\prime}) = j_0 \left( k \| \bm{r}-\bm{r}^{\prime} \| \right).
\label{eq:ker}
\end{align}
This function is equivalent to the spatial covariance function of diffused sound fields. Thus, the kernel ridge regression with the constraint of the Helmholtz equation is achieved by using the kernel function defined in \eqref{eq:ker-dir} or \eqref{eq:ker}. Figure~\ref{fig:sf_est_kernel} is an example of the experimental comparison of the physics-constrained kernel function \eqref{eq:ker} and the Gaussian kernel function.

The kernel regression described above is applied to interpolate the pressure distribution from discrete pressure measurements. This technique can be generalized to the cases of using directional microphones and estimating expansion coefficients of spherical wave functions~\cite{Ueno:IEEE_J_SP2021}, which is regarded as an infinite-dimensional generalization of the finite-dimensional basis expansion method~\cite{Laborie:AESconv2003}.

\subsection{Regression by neural networks incorporating governing PDEs}
\label{sec:nn}

\begin{figure*}[!t]
\centering
\includegraphics[width=5.0in,clip]{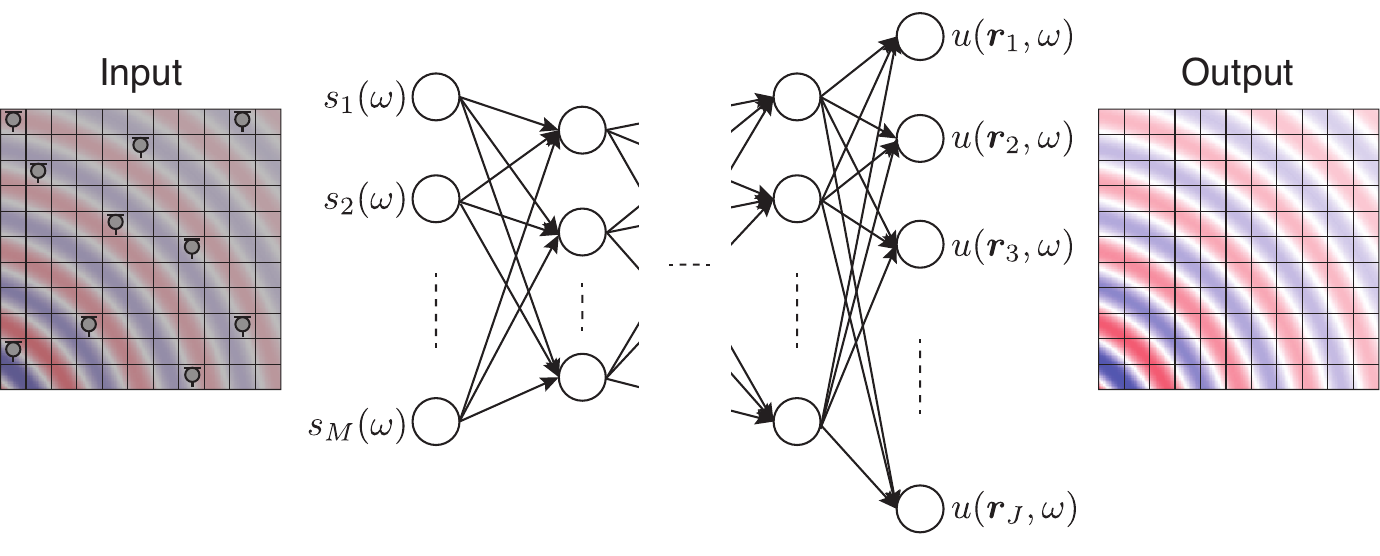}
\caption{Feedforward neural network for sound field estimation in the frequency domain. The input is the observation $\bm{s}$ and the target output is the discretized value of $u$. The network weights are optimized by using a pair of datasets.}
\label{fig:ffnn}
\end{figure*}

When applying neural networks to regression problems, the target output is typically a discretized value, which is denoted as $\bm{t} \in \mathbb{R}^J$, considering the real-valued function $f$. In a simple manner, $\bm{t}$ is defined as a discretization of the domain into $\{\bm{x}_j\}_{j=1}^J$ with $J$ ($>I$) sampling points, i.e., $t_j = f(\bm{x}_j)$. The neural network with input $\bm{y}\in\mathbb{R}^I$ and output $\bm{g}(\bm{y})\in \mathbb{R}^J$, as shown in Fig.~\ref{fig:ffnn}, is trained using a pair of datasets for various sampling points and function values $\{(\bm{y}_d, \bm{t}_d)\}_{d=1}^D$, where the subscript $d$ is the index of the data and $D$ is the number of training data. There are various possibilities of the network architecture, but its parameters are collectively denoted as $\bm{\theta}_{\mathrm{NN}}$. The loss function used to train the network, i.e., to obtain $\bm{\theta}_{\mathrm{NN}}$, is, for example, the squared error written as
\begin{align}
 \mathcal{J}(\bm{\theta}_{\mathrm{NN}}) = \sum_{d=1}^D \| \bm{t}_d - \bm{g}(\bm{y}_d;\bm{\theta}_{\mathrm{NN}})\|^2.
\end{align}
The parameters $\bm{\theta}_{\mathrm{NN}}$ are optimized basically by minimizing $\mathcal{J}$ using steepest-descent-based algorithms~\cite{Goodfellow:DL}. In the inference, the output for the unknown observation $\bm{y}$, i.e., $\bm{g}(\bm{y})$, is computed by forward propagation, which is regarded as a discretized estimate of $f$. When estimating complex-valued functions, a simple strategy is to treat real and imaginary values separately. Other techniques are, for example, separating the complex values into magnitudes and phases and using complex-valued neural networks~\cite{Hirose:CVNN}.

It is not straightforward to incorporate physical properties into the neural-network-based regression because the output is a discretized value. One of the techniques to constrain the estimate to satisfy the Helmholtz equation is to set the expansion coefficients of element solutions introduced in Section~\ref{sec:basis-exp} to the target output~\cite{Karakonstantis:JASA2023,Lobato:JASA2024}. By using a pair of datasets $\{(\bm{y}_d, \bm{\gamma}_d)\}_{d=1}^D$, we can train the network for predicting expansion coefficients. By using the estimated $\bm{\gamma}$, we can obtain the continuous function $f$ satisfying the Helmholtz equation. This technique is referred to as the \textit{physics-constrained neural networks (PCNNs)}.

\subsection{PINNs based on implicit neural representation}
\label{sec:pinn}

\begin{figure*}[!t]
\centering
\includegraphics[width=6.0in,clip]{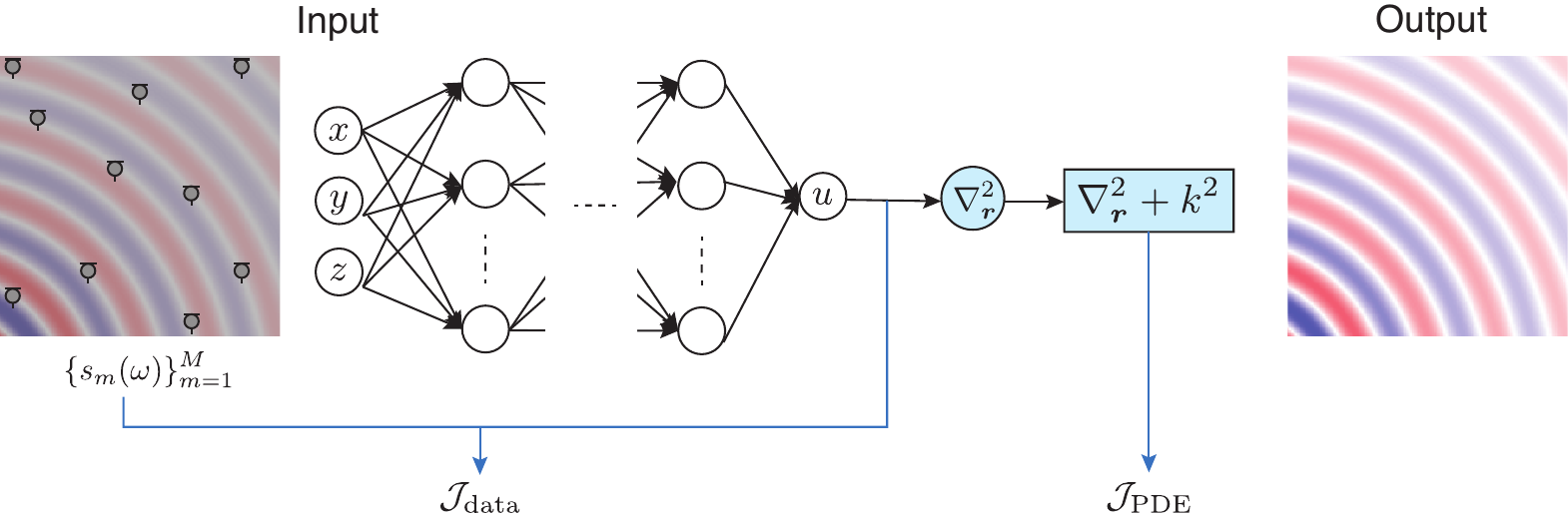}
\caption{PINN using implicit neural representation for sound field estimation in the frequency domain. The input is the positional coordinates $\bm{r}=(x,y,z)$ and the target output is $u(\bm{r})$. The network weights are optimized by using the observation $\bm{y}$. In PINNs, the loss function is composed of the data loss $\mathcal{J}_{\mathrm{data}}$ and PDE loss $\mathcal{J}_{\mathrm{PDE}}$ to penalize the deviation from the governing PDE.}
\label{fig:pinn}
\end{figure*}

PINNs were first proposed to solve forward and inverse problems containing PDEs by using the implicit neural representation~\cite{Raissi:CompPhys2019,Karniadakis:NatRevPhus2021}. In the implicit neural representation, neural networks (typically, multilayer perceptions) are used to implicitly represent a continuous function $f$. The input and output of a neural network are the argument of $f$, i.e., $\bm{x}\in\mathbb{R}^P$, and the scalar function value $g(\bm{x};\bm{\theta}_{\mathrm{NN}})\in\mathbb{R}$ approximating $f(\bm{x})$, respectively (see Fig.~\ref{fig:pinn}). The training data in this scenario are the pair of sampling points and function values $\{(\bm{x}_i, y_i)\}_{i=1}^I$. The cost function is, for example, the squared error written as
\begin{align}
 \mathcal{J}_{\mathrm{INR}}(\bm{\theta}_{\mathrm{NN}}) = \sum_{i=1}^I |y_i - g(\bm{x}_i; \bm{\theta}_{\mathrm{NN}})|^2.
\end{align}
Therefore, it is unnecessary to discretize the target output or prepare a pair of datasets because the observation data are regarded as the training data in this context.

A benefit of the implicit neural representation is that some constraints on $g$ including its (partial) derivatives can be included in the loss function:
\begin{align}
 \mathcal{J}_{\mathrm{INR}}(\bm{\theta}_{\mathrm{NN}}) = \sum_{i=1}^I |y_i - g(\bm{x}_i; \bm{\theta}_{\mathrm{NN}})|^2 + \epsilon \sum_{n=1}^N \left| H \left(g(\bm{x}_n), \nabla_{\bm{x}} g(\bm{x}_n), \nabla_{\bm{x}}^2 g(\bm{x}_n), \ldots  \right) \right|^2
\end{align}
with the positive constant $\epsilon$ and given constraints at $\{\bm{x}_n\}_{n=1}^N$
\begin{align}
 H \left(g(\bm{x}_n), \nabla_{\bm{x}} g(\bm{x}_n), \nabla_{\bm{x}}^2 g(\bm{x}_n), \ldots \right) = 0.
\end{align}
Here, $\nabla_{\bm{x}}$ represents the gradient with respect to $\bm{x}$. The evaluation points of the constraints $\{\bm{x}_n\}_{n=1}^N$ can be determined independently from the sampling points $\{\bm{x}_i\}_{i=1}^I$. The loss function including derivatives is usually calculated using automatic differentiation~\cite{Baydin:JMLR2018}.

\begin{figure*}
    \centering
    \includegraphics[width=6in,clip]{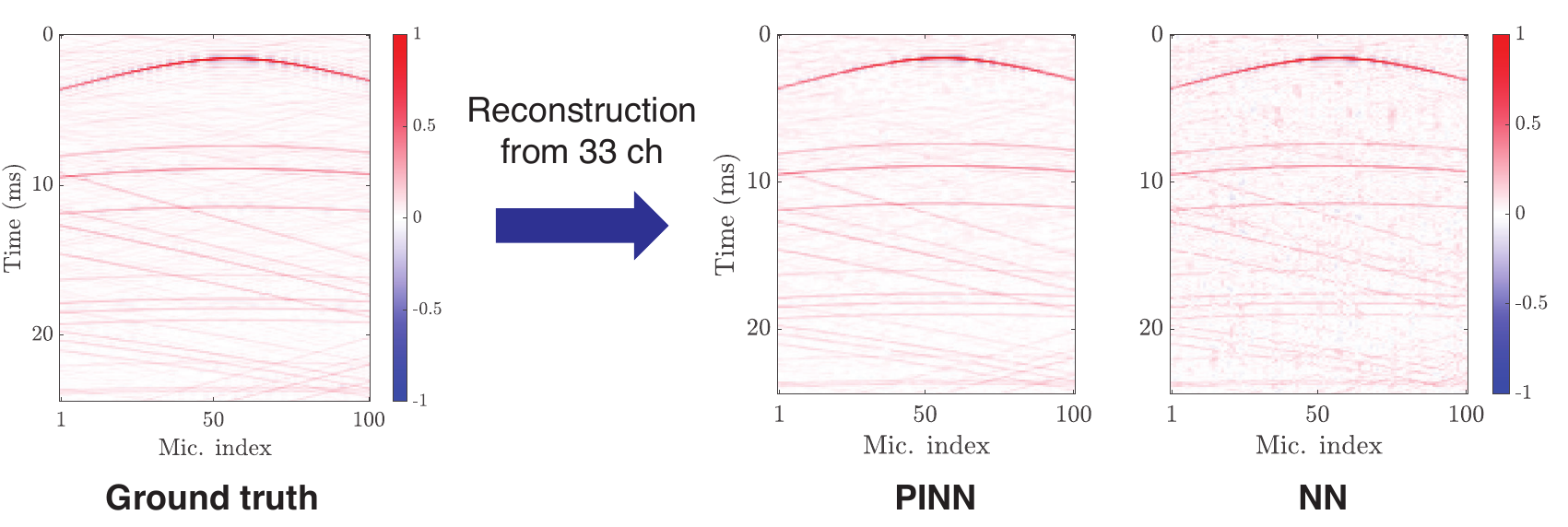}
    \caption{RIRs (simulated shoebox-shaped room of $T_{60}=500~\mathrm{ms}$) measured by a uniform linear array of $100$ microphones are reconstructed using only $33$ randomly selected channels. PINN~\cite{Olivieri:EURASIP2024} and a plain neural network (NN) are compared. The addition of the physics loss improved the reconstruction accuracy of RIRs. Figures are taken from \cite{Pezzoli:ForumAcusticum2023} in which the method proposed in \cite{Olivieri:EURASIP2024} has initially appeared, and further details of the experimental setting are presented.}
    \label{fig:SIREN}
\end{figure*}

Therefore, the physical properties represented by a PDE can be embedded into the loss function to infer the function approximately satisfying the governing PDE. When considering the sound field estimation in the frequency domain, as shown in Fig.~\ref{fig:pinn}, the loss function to train the neural network $g$ with the input $\bm{r}\in\Omega$ is expressed as
\begin{align}
 \mathcal{J}_{\mathrm{PINN}}(\bm{\theta}_{\mathrm{NN}}) = \mathcal{J}_{\mathrm{data}} +  \epsilon \mathcal{J}_{\mathrm{PDE}},
\label{eq:pinn_loss}
\end{align}
where 
\begin{align}
 & \mathcal{J}_{\mathrm{data}} = \sum_{m=1}^M |s_m(\omega) - g(\bm{r}_m; \bm{\theta}_{\mathrm{NN}})|^2, \\
 & \mathcal{J}_{\mathrm{PDE}} = \sum_{n=1}^N |(\nabla_{\bm{r}}^2 + k^2)g(\bm{r}_n; \bm{\theta}_{\mathrm{NN}})|^2.
\end{align}
Thus, $g$ is trained to minimize the deviation from the Helmholtz equation as well as the observations at the sampling points $\{\bm{r}_m\}_{m=1}^M$. Since this technique is based on an implicit neural representation, the observed signal $\{s_m(\omega)\}_{m=1}^M$ corresponds to the training data for the neural network. In this sense, PINNs can be classified as an unsupervised approach. 
Figure~\ref{fig:SIREN} shows an experimental example of RIR reconstruction using a neural network and the same architecture trained using the PINN loss \eqref{eq:pinn_loss}. 
It can be seen in Fig.~\ref{fig:SIREN} that the neural network can estimate the RIRs, although they contain noise and incoherent wavefronts in the missing channels.
In contrast, the result of the PINN is more accurate with reduced noise and coherent wavefronts owing to the adoption of the PDE loss.

To extract the properties of an acoustic environment by using a set of training data, i.e., supervised learning for sound field estimation, the general neural networks in Section~\ref{sec:nn} can be applied~\cite{Lluis:JASA2020,Pezzoli:Sensors2022,Fernandez-Grande:JASA2023,Miotello:ICASSP2024}. Since the output of the neural network is a discretized value, the evaluation of the deviation from the governing PDE is not straightforward. A simple approach is difference approximation by finely discretizing the domain to compute the PDE loss~\cite{Zhao:EAAI2023}. Another strategy is the use of general interpolation techniques to obtain a continuous function from the output of the neural networks~\cite{Shigemi:IWAENC2022}; thus, the PDE loss is computed by using the interpolated function.

\section{Current studies of sound field estimation based on PIML}
\label{sec:sota}

Current sound field estimation methods based on PIML are essentially based on the techniques introduced in Section~\ref{sec:emb-phys}. The linear ridge regression based on finite-dimensional expansion into basis functions has been thoroughly investigated in the literature. In particular, the spherical wave function expansion has been used in spatial audio applications because of its compatibility with spherical microphone arrays~\cite{Poletti:J_AES_2005,Rafaely:FundSphArrayProc}. The kernel ridge regression (or infinite-dimensional expansion) with the constraint of the Helmholtz equation can be regarded as a generalization of linear-ridge-regression-based methods. Since it is applicable to the estimation in a region of an arbitrary shape with arbitrarily placed microphones and the number of parameters to be manually tuned can be reduced, kernel-ridge-regression-based methods are generally simpler and more useful than the linear-ridge-regression-based methods. One of the advantages of these methods is a linear estimator in the frequency domain, which means that the estimation is performed by the convolution of an FIR filter in the time domain. This property is highly important in real-time applications such as active noise control. 

Sparsity-based methods have also been intensively investigated over the past ten or more years both in the frequency and time domains~\cite{Bertin:CSAbook2015,Antonello:IEEE_ACM_J_ASLP2017,Murata:IEEE_J_SP2018}. Sparsity is usually imposed on expansion coefficients of plane waves \eqref{eq:pw-exp} and equivalent sources \eqref{eq:esd}. Since the sound field is represented by a linear combination of sparse plane waves or equivalent sources, it is guaranteed that the interpolated function satisfies the Helmholtz equation. These techniques are effective when the sound field to be estimated fits the sparsity assumption, such as a less reverberant sound field. However, iterative algorithms are required to obtain a solution; thus, the estimator is nonlinear, and STFT-based processing is necessary in practice. The sparse plane wave model is also used in hierarchical-Bayesian-based Gaussian process regression, which can be regarded as a generalization of the kernel ridge regression~\cite{Nozal:JASA2021}. Then, the inference process becomes a linear operation. 

Neural-network-based sound field estimation methods have gained interest in recent years, which makes the inference more adaptive to the environment owing to the high representational power of neural networks~\cite{Lluis:JASA2020,Pezzoli:Sensors2022,Luo:NIPS2022,Fernandez-Grande:JASA2023,Liang:EURASIP2024,Miotello:ICASSP2024}. Physical properties have recently been incorporated into neural networks to mitigate physically unnatural distortions even with a small number of microphones. PINNs for the sound field estimation problem are investigated both in the frequency~\cite{Ribeiro:TechRxiv2023,Chen:APSIPA2023,Ma:arxiv2024} and time domains~\cite{Olivieri:EURASIP2024,Karakonstantis:JASA2024}. A major difference from the methods based on basis expansions and kernels is that the estimated function does not strictly satisfy the governing PDEs, i.e., the Helmholtz and wave equations, because its constraint is imposed by the loss function $\mathcal{J}_{\mathrm{PDE}}$ as in \eqref{eq:pinn_loss}. The governing PDEs are strictly satisfied by using PCNN, i.e., setting the target output as expansion coefficients of the basis expansions~\cite{Karakonstantis:JASA2023,Lobato:JASA2024}; however, this approach can sacrifice the high representational power of neural networks by strongly restricting the solution space. Whereas the methods based on implicit neural representation are regarded as unsupervised learning, supervised-learning-based methods may have the capability to capture properties of acoustic environments and high accuracy even when the number of available microphones is particularly small. In the physics-informed convolutional neural network~\cite{Shigemi:IWAENC2022}, the PDE loss is computed by using bicubic spline interpolation. In \cite{Oliveri:Sensors2021}, a loss function based on the boundary integral equation of the Helmholtz equation, called the Kirchhoff--Helmholtz integral equation~\cite{Williams:FourierAcoust}, is proposed in the context of acoustic imaging using a planar microphone array, which is also regarded as a supervised-learning approach. 

\begin{figure*}[!t]
\centering
\includegraphics[width=6.0in,clip]{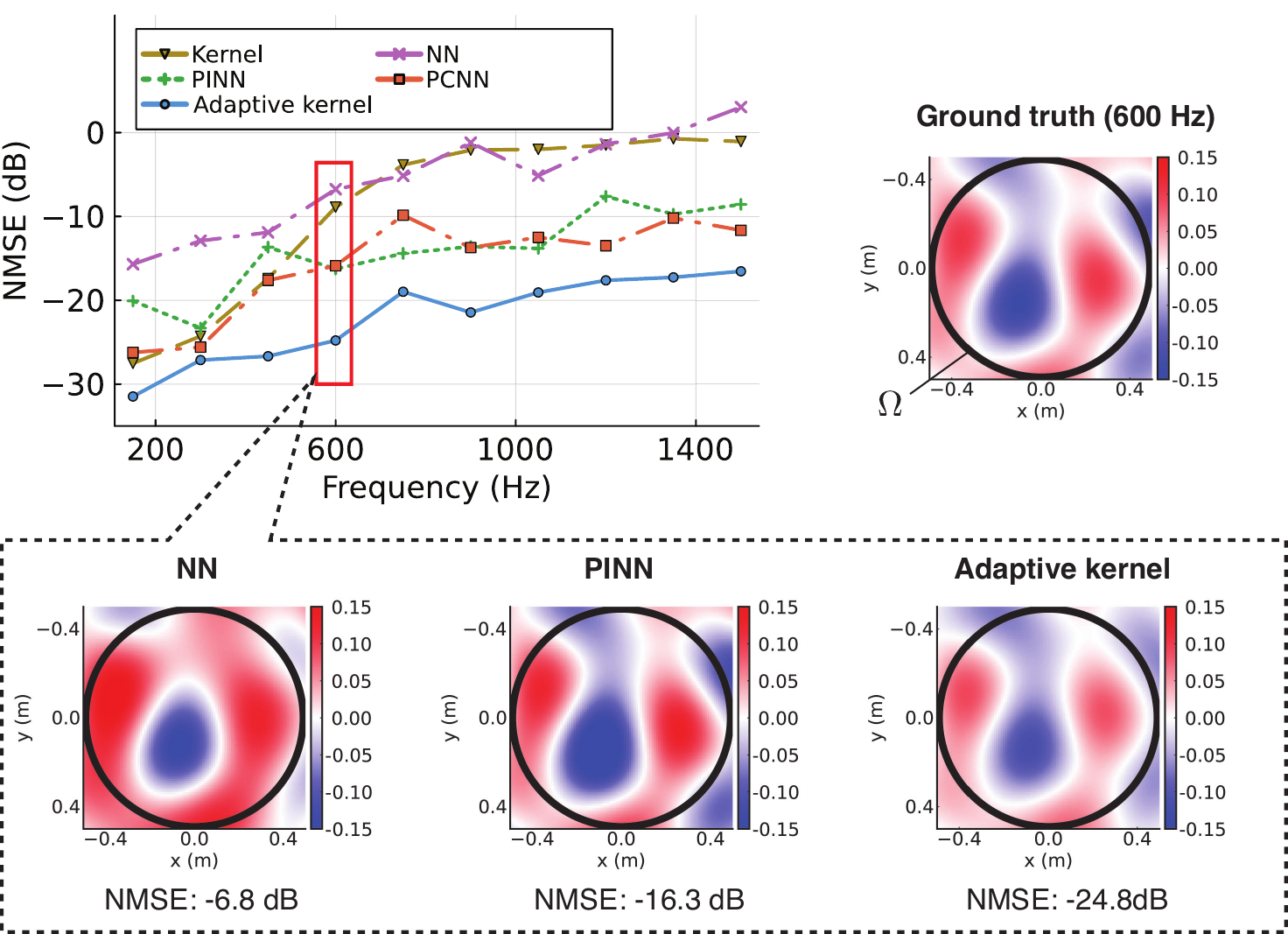}
\caption{Normalized mean square error (NMSE) and reconstructed pressure distribution inside a sphere of the target region in the frequency domain. A single point source was located outside the target region, and RIRs in a shoebox-shaped room were simulated by the image source method so that the reverberation time $T_{60}$ becomes $400~\mathrm{ms}$. $M=41$ microphones were placed on two spherical surfaces with radii of $0.50$ and $0.49~\mathrm{m}$. The kernel method using \eqref{eq:ker}, the neural network without any physics (NN), PINN, PCNN with plane wave decomposition, and the adaptive kernel method proposed in \cite{Ribeiro:ICASSP2023,Ribeiro:TechRxiv2023} are compared. The number of evaluation points of the PDE loss in PINN was $N=56$. The architectures of NN, PINN, PCNN, and the adaptive kernel are basically the same (fully connected layers with 3 entries for input, 3 hidden layers with 8 neurons for the first, 8 neurons for neural ordinary differential equation (NODE) layer~\cite{Chen:NIPS2018}, and 5 neurons for the third hidden layer, and 1 output neuron. The rowdy activation function~\cite{Jagtap:Neurocomput2022} is used for the first three layers and ReLU for the last layer on PCNN and adaptive kernel to guarantee nonnegativity.). Further details of the experimental setting are presented in \cite{Ribeiro:TechRxiv2023}.}
\label{fig:nmse_current}
\end{figure*}

The inference process of neural networks involves multiple linear and nonlinear transformations; therefore, if a network is designed in the frequency domain, it cannot be implemented as an FIR filter in the time domain. The kernel-ridge-regression-based method using the kernel function adapted to acoustic environments using neural networks enables the linear operation of inference while preserving the constraint of the Helmholtz equation and the high flexibility of neural networks. In \cite{Ribeiro:ICASSP2023,Ribeiro:TechRxiv2023}, the directional weight $w$ in \eqref{eq:ker-ip} is separated into the directed and residual components. The directed component is modeled by a weighted sum of von Mises--Fisher distributions \eqref{eq:w_dir} and the residual component is represented by a neural network to separately represent highly directive sparse sound waves and spatially complex late reverberation. The parameters of these components are jointly optimized by a steepest-descent-based algorithm. Experimental results in the frequency domain are demonstrated in Fig.~\ref{fig:nmse_current}, where the kernel method using \eqref{eq:ker}, a plain neural network, PINN, PCNN with plane wave decomposition, and the adaptive kernel method proposed in \cite{Ribeiro:ICASSP2023,Ribeiro:TechRxiv2023} are compared by numerical simulation. It can be observed that the adaptive kernel method performs well over the investigated frequency range.

As described above, various attempts have been made to incorporate physical properties into function interpolation and regression techniques for sound field estimation. The expansion representation introduced in Section~\ref{sec:basis-exp} leads to the closed-form linear estimator in the frequency domain based on the linear and kernel ridge regression techniques. In exchange for their simple implementation and small computational cost, these techniques do not have sufficient adaptability to the acoustic environment because of their fixed estimator. Methods for adaptive estimators, e.g., using sparse optimization and kernel learning techniques, have also been studied. Recent neural-network-based methods having both high flexibility and physical validity have been investigated for both supervised- and unsupervised-learning approaches. Many of them make the estimator nonlinear; thus, the inference process for each time or time--frequency data becomes the forward propagation of the neural network for the supervised methods and both the training and inference of the neural network for the unsupervised methods; this process requires a high computational load. However, several methods integrating the expansion representation and neural networks enable estimators to be linear, which will be suitable for real-time applications. The current PIML-based sound field estimation methods using neural networks are compared in Table~\ref{tbl:comp}.

\begin{table*}[t]
 \caption{Comparison of current sound field estimation methods based on PIML using neural networks with training data required or not required, linear or nonlinear estimator, time or frequency domain, and penalized or constrained physical property.}
 \label{tbl:comp}
 \centering
 \begin{tabular}[t]{c|c|c|c|c}
  \hline
  & Supervised/unsupervised & Estimator & Domain & Physical property \\
  \hline \hline
  %Oliverli~et~al.~2021~\cite{Oliveri:Sensors2021} & Supervised & Nonlinear & Frequency & Penalized \\
  Shigemi~et~al.~2022~\cite{Shigemi:IWAENC2022} & Supervised & Nonlinear & Frequency & Penalized \\
  \begin{tabular}{c} Karakonstantis~et~al.~2023~\cite{Karakonstantis:JASA2023}\\ and Lobato~et~al.~2024~\cite{Lobato:JASA2024} \end{tabular} & Supervised & Linear & Frequency & Constrained \\
  \begin{tabular}{c} Chen~et~al.~2023~\cite{Chen:APSIPA2023}\\ and Ma~et~al.~\cite{Ma:arxiv2024} \end{tabular} & Unsupervised & Nonlinear & Frequency & Penalized \\
  \begin{tabular}{c} Olivieri~et~al.~2024~\cite{Olivieri:EURASIP2024}\\ and Karakonstantis~et~al.~2024~\cite{Karakonstantis:JASA2024} \end{tabular} & Unsupervised & Nonlinear & Time & Penalized \\
  Ribeiro~et~al.~2023~\cite{Ribeiro:ICASSP2023,Ribeiro:TechRxiv2023} & Unsupervised & Linear & Frequency & Constrained \\
  \hline
 \end{tabular}
\end{table*}

\section{Outlook}
\label{sec:outlook}

PIML for sound field estimation is a rapidly growing research field and has successfully been applied to some application scenarios. However, the current methods still have some limitations. In particular, the major problems that should be resolved are 1) the preparation of training data, 2) the mismatch between training and test data, and 3) the neural network architecture design.

The methods using only the observation data, i.e., unsupervised methods, are useful when the acoustic environment to be measured is unpredictable. Supervised methods have the potential to extract properties of an acoustic environment by using a set of training data and to improve estimation accuracy when the number of microphones is particularly small. However, the training data should be a set of sound field data with high spatial (or spatiotemporal) resolution. Since the acoustic environment can have large variations of, e.g., source positions, directivity, and reflections, the first problem is how to prepare the training data, even though physical properties could help prevent overfitting in the case of a small number of training data. Currently, it is considered impractical to collect all the variations of sound field data by actual measurements. Therefore, it will be necessary to integrate multiple training datasets measured in different environments and also include some simulated data. Thus, the second problem of the mismatch between training and test data arises. The measurement systems for acoustic fields could have some variations. Even for the wave-based acoustic simulation techniques requiring a high computational load, there are some deviations between real and simulated data. Furthermore, the governing PDEs still include given model parameters, e.g., sound speed, which can be different from those in the training data. It is still unclear how these deviations can affect the estimation performance and how to mitigate those effects. The third problem is generally common to neural-network-based techniques, but there is currently no clear methodology for architectural design. Therefore, at present, architectures must be designed by trial and error depending on the task and data. 

In addition to the problems listed above, there are several other unsolved issues. First, the target region of estimation is usually assumed not to include any sound sources or scattering objects. This assumption allows us to represent the interior sound field by basis functions introduced in Section~\ref{sec:basis-exp}. However, in some applications, the target region is not necessarily source-free; thus, the governing PDE of the sound field to be estimated can become inhomogeneous. Several methods have been proposed for estimating a sound field including sound sources or scatterers. Furthermore, the methods using the PDE loss, such as PINN, can also be applied in this setting. However, further investigations of these methods in practical applications are still necessary.

Second, the estimation accuracy of the sound field estimation methods is highly dependent on the number of microphones and their placement. It is often necessary to cover the entire target region with the smallest possible number of microphones. On the basis of boundary integral equations, such as Kirchhoff--Helmholtz integrals, a sound field inside a source-free interior region can be predicted by using pressure and its normal derivative on its surface. However, the interior sound field cannot be uniquely determined at some frequencies solely from the pressure measurements on the boundary surface, which is called the \textit{forbidden frequency problem}~\cite{Williams:FourierAcoust}. Although it is necessary to place microphones inside the target region, its optimal placement is not straightforward, especially for broadband measurements. Several techniques for the optimal placement of microphones for sound field estimation have been proposed; however, their effectiveness has not been sufficiently evaluated in practical applications. It may also be possible to incorporate auditory properties and/or multimodal information to reduce the number of microphones. 

The PIML-based sound field estimation techniques, particularly linear-ridge-regression and kernel-ridge-regression-based methods, have already been applied to various applications, e.g., the spatial audio recording for binaural reproduction~\cite{Iijima:JASA_J_2021,Birnie:IEEE_J_ASLP2021}, spatial interpolation of HRTFs~\cite{Duraiswami:ICASSP2004}, and spatial active noise control~\cite{Zhang:IEEE_ACM_J_ASLP2018,Koyama:IEEE_ACM_J_ASLP2021}, owing to their simple implementation and small computational cost. In binaural reproduction, the signals reaching both ears, when a listener is present in the target region, are reproduced by estimating the expansion coefficients of the sound field at the listening position. The HRTFs are also required in the binaural reproduction, but measuring them for each listener is costly. The interpolation of HRTFs can be formulated as an exterior sound field estimation problem based on reciprocity. Active noise control in a spatial region requires estimating a sound field of primary noise using microphones in a target region and synthesizing an anti-noise sound field using loudspeakers in real time. On the other hand, applications of neural-network-based methods are still limited to offline applications, e.g., acoustic imaging, owing to the above-mentioned challenges as well as the issue of relatively high computational cost. It is expected that these methods will also be applied to various applications in the future because of their high flexibility and interpolation capability.

\section{Conclusion}
PIML-based sound field estimation methods are overviewed. Our focus was the interior problem, but several techniques can also be applied to the exterior problem and estimation in a region containing some sources or scattering objects. There exist several techniques that can be used to embed physical properties in interpolation techniques. In particular, the governing PDE of the sound field is considered useful prior information. Neural-network-based methods, which have been successfully applied in various domains, are expected to perform well for sound field estimation owing to their high expressive power. How to integrate physical prior knowledge with the high expressive power of neural networks will continue to be one of the most important topics in sound field estimation and its applications. Although there are still many unresolved issues, such as the preparation of training data, we expect further development of sound field estimation methods and their applications in various applied technologies in the future.

%\section*{Acknowledgments}
%This should be a simple paragraph before the References to thank those individuals and institutions who have supported your work on this article.

%{\appendix[Proof of the Zonklar Equations]
%Use $\backslash${\tt{appendix}} if you have a single appendix:
%Do not use $\backslash${\tt{section}} anymore after $\backslash${\tt{appendix}}, only $\backslash${\tt{section*}}.
%If you have multiple appendixes use $\backslash${\tt{appendices}} then use $\backslash${\tt{section}} to start each appendix.
%You must declare a $\backslash${\tt{section}} before using any $\backslash${\tt{subsection}} or using $\backslash${\tt{label}} ($\backslash${\tt{appendices}} by itself
% starts a section numbered zero.)}


%\begin{thebibliography}{1}
%\bibliographystyle{IEEEtran}
%\end{thebibliography}

\bibliographystyle{IEEEtran}
\bibliography{str_def_abrv,koyama_en,refs}

%\newpage

\begin{IEEEbiographynophoto}{Shoichi Koyama}
(koyama.shoichi@ieee.org) received the B.E., M.S, and Ph.D. degrees from the University of Tokyo, Tokyo, Japan, in 2007, 2009, and 2014, respectively. He is currently an Associate Professor at the National Institute of Informatics (NII), Tokyo, Japan. Prior to joining NII, he was a Researcher at Nippon Telegraph and Telephone Corporation (2009--2014), and Research Associate (2014--2018) and Lecturer (2018--2023) at the University of Tokyo, Tokyo, Japan. He was also a Visiting Researcher at Paris Diderot University (Paris 7), Institut Langevin, Paris, France (2016--2018), and a Visiting Associate Professor at Research Institute of Electrical Communication, Tohoku University, Miyagi, Japan (2020--2023). His research interests include audio signal processing/machine learning, acoustic inverse problems, and spatial audio. 
\end{IEEEbiographynophoto}

\begin{IEEEbiographynophoto}{Juliano G. C. Ribeiro}
Juliano G. C. Ribeiro received his B.E. degree in electronic engineering from the Instituto Militar de Engenharia (IME), Rio de Janeiro, Brazil, in 2017, and the M.S. and Ph.D. degrees in information science and technology from the University of Tokyo, Tokyo, Japan, in 2021 and 2024, respectively. Currently a Research Engineer with Yamaha corporation. His research interests include spatial audio, acoustic signal processing and physics-informed machine learning.
\end{IEEEbiographynophoto}

\begin{IEEEbiographynophoto}{Tomohiko Nakamura}
(tomohiko.nakamura.jp@ieee.org) received B.S., M.S., and Ph.D. degrees from the University of Tokyo, Japan, in 2011, 2013, and 2016, respectively. He joined SECOM Intelligent Systems Laboratory as a researcher in 2016 and moved to the University of Tokyo as Project Research Associate in 2019. He is currently a Senior Researcher with the National Institute of Advanced Industrial Science and Technology (AIST). His research interests include signal-processing-inspired deep learning, audio signal processing, and music signal processing.
\end{IEEEbiographynophoto}

\begin{IEEEbiographynophoto}{Natsuki Ueno}
(natsuki.ueno@ieee.org) received the B.E. degree in engineering from Kyoto University, Kyoto, Japan, in 2016, and the M.S. and Ph.D. degrees in information science and technology from the University of Tokyo, Tokyo, Japan, in 2018 and 2021, respectively. 
% He is currently a Project Assistant Professor at Tokyo Metropolitan University, Tokyo, Japan. 
He is currently an Associate professor at Kumamoto University, Kumamoto, Japan. 
His research interests include spatial audio and acoustic signal processing.
\end{IEEEbiographynophoto}

\begin{IEEEbiographynophoto}{Mirco Pezzoli}
(mirco.pezzoli@polimi.it) received the M.S. degree (cum laude), in 2017, in computer engineering from the Politecnico di Milano, Italy. In 2021 he received the Ph.D. degree in information engineering at Politecnico di Milano. After two years as a Postdoctoral Researcher, he joined the Department of Electronics, Information and Bioengineering of the Politecnico di Milano as Junior Assistant Professor. His main research interests are multichannel audio signal processing, sound field reconstruction and musical acoustics.
\end{IEEEbiographynophoto}

%\bf{If you include a photo:}\vspace{-33pt}
%\begin{IEEEbiography}[{\includegraphics[width=1in,height=1.25in,clip,keepaspectratio]{fig1}}]{Michael Shell}
%Use $\backslash${\tt{begin\{IEEEbiography\}}} and then for the 1st argument use $\backslash${\tt{includegraphics}} to declare and link the author photo.
%Use the author name as the 3rd argument followed by the biography text.
%\end{IEEEbiography}



\vfill

\end{document}